\documentclass[aps,prd,reprint,nofootinbib,twocolumn]{revtex4-1}

\usepackage[english]{babel}
\usepackage{amsmath}
\usepackage{amsfonts}
\usepackage{graphicx}
\usepackage[colorlinks=true, allcolors=blue]{hyperref}
\usepackage{cancel}
\usepackage{tikz}

\usepackage[export]{adjustbox}
\hypersetup{breaklinks=true}

\usepackage{comment}
\usepackage{MnSymbol}
\usepackage{graphicx}

\def\eqrev#1{\textcolor{red}{#1}}

\newcommand{\trphi}{${\rm Tr}(\Phi^3)$~}

\newcommand{\VSI}{Van Swinderen Institute for Particle Physics and Gravity,\\ University of Groningen,
Nijenborgh 3, 9747 AG Groningen, The Netherlands}

\newcommand{\mac}{Department of Physics and Astronomy,
Macalester College,
Saint Paul MN 55105-1899, U.S.A.}

\begin{document}
\title{Hidden Zeros in Exceptional Field Theories from Double Copy}
\author{Yang Li$^{1}$\footnote{y.l.li@rug.nl}, Diederik Roest$^{1}$\footnote{d.roest@rug.nl}, Tonnis ter Veldhuis$^{2}$\footnote{terveldhuis@macalester.edu}}
\affiliation{$^{1}$\VSI \\ 
$^{2}$\mac}

\begin{abstract}
    It was recently discovered by Arkani-Hamed et al and Cao et al that the colour-ordered scattering amplitudes of Tr$(\Phi^3)$, the non-linear sigma model and Yang-Mills-scalar vanish at specific loci. We build on this observation and demonstrate that, beyond colour ordering, scattering amplitudes can display higher-order hidden zeros. A first example are the flavour-paired amplitudes of  gravity-coupled scalars, as a double copy of Yang-Mills-scalar. The other two cases are Dirac-Born-Infeld and the special Galileon, which are the natural generalisations of the non-linear sigma model with higher orders of the Adler zero. We demonstrate that the amplitudes of these theories all factorize into lower-point objects in the near-zero limit, and discuss their interpretation. Finally, we comment on the general picture of hidden zeros and prove their relation under the double copy.
\end{abstract}

\maketitle

\section{Introduction}

\noindent
Positive geometry has been shown to encode rich properties of scattering amplitudes that are hidden from the Lagrangian prospective. For instance, the canonical form \cite{Arkani-Hamed:2017tmz} of the kinematic associahedron \cite{Arkani-Hamed:2017mur} was found to be directly related to bi-adjoint scalar (BAS) amplitudes. Furthermore, its $\alpha'$-generalization  gives rise to amplitudes of stringy nature \cite{Arkani-Hamed:2019mrd}. Much more recently, it was found that the amplitudes of \trphi theory (that are contained in those of BAS) have a surprising property with a beautiful geometric structure: at all-loop and of any multiplicity, these follow from compact geometric rules on surfaces \cite{Arkani-Hamed:2023lbd,Arkani-Hamed:2023mvg}. This framework of surfaceology has led to a series of interesting follow-up works \cite{Arkani-Hamed:2023jry,Arkani-Hamed:2024nhp,Arkani-Hamed:2024vna,Arkani-Hamed:2024yvu}. 

The tree-level amplitudes of \trphi vanish on specific hypersurfaces in the space of Mandelstam variables \cite{Arkani-Hamed:2023swr} (as also found much earlier in string theory \cite{DAdda:1971wcy}). These loci naturally arise as 'causal diamonds' in a two-dimensional kinematic mesh \cite{Arkani-Hamed:2019vag}, that is defined with respect to a specific colour ordering (and applies to the corresponding partial amplitude). On top of this, the loci give rise to new factorisation channels into lower-point amplitudes of the same theory.

These minor miracles carry over to more intricate and interesting theories. Two of these were identified in \cite{Arkani-Hamed:2023swr} and involve colour ordering as well. The first is the scattering of pions, as described by the chiral non-linear sigma model (NLSM). The non-linear symmetry of these famously gives rise to the Adler zero in the soft limit \cite{Adler:1964um}. Interestingly, the zeros of Arkani-Hamed et al include and extend this soft-limit behaviour. The second theory describes external scalars interacting via gluon exchange and a quartic interaction, referred to as Yang-Mills-scalar (YMS) \cite{Cachazo:2014xea}, whose factorisation has been discussed in \cite{Cao:2024gln,Cao:2024qpp}.

In this work, we will identify three theories that naturally generalise these results. The first two are higher-order variations of the Adler zero: the amplitudes of Dirac-Born Infeld (DBI) and the special Galileon (SG) vanish in the soft limit as $p^\sigma$ with $\sigma=2,3$, respectively  \cite{Cheung:2016drk}. Similar to the Adler zero, we will demonstrate that their amplitudes vanish at multiple loci that include and extend these soft-limit behaviours. The third theory describes flavoured scalars minimally coupled to Einstein gravity, i.e.~gravity-coupled scalars (GCS).

Together, these six theories naturally span the set of scalar field theories that are related via the double copy procedure \cite{Bern:2008qj,Bern:2010ue,Bern:2019prr,Bern:2022wqg}. One can think of \trphi as the zeroth copy, YMS and the NLSM as single copies, completed by GCS, DBI and the special Galileon as double copies. We use the KLT formulation of these relations to demonstrate how the single-copy zeros carry over to the double copies, thus completing the web of exceptional scalar field theories with zeros in their amplitudes. Crucial input for these relations are the hidden zeros of YMS, which we have explicitly checked up to 8-point and which we expect to hold up to arbitrary multiplicity. \\

\noindent
{\bf Note added - } Upon completion of this manuscript, the preprints \cite{Cao:2024gln, Bartsch:2024amu} appeared whose results overlap in parts with ours.

\section{Colour ordering \& single zeros}
\label{sec: color order zeros}

\noindent
We first recap the results of \cite{Arkani-Hamed:2023swr} on the structure of hidden zeros in the colour-ordered partial amplitudes of Tr$(\Phi^3)$ and the NLSM, followed by YMS \cite{Cao:2024gln, Cao:2024qpp}. \\

\noindent
{\bf Tr$(\Phi^3)$ theory -} 
Starting at four-point (4-pt) and requiring the colour-ordered partial amplitude to vanish at $u=0$, there are two poles available for exchange interactions: one can include contributions via the $s$- and $t$-channel. At lowest order in derivatives, the unique possibility is 
 \begin{align}
  A^{\Phi^3}_4 \sim \frac1s + \frac1t \sim \frac{u}{st} \,. \label{A4-Phi3}
 \end{align}
One can identify this as the partial amplitude (with alphabetic colour ordering) of the Tr$(\Phi^3)$ theory.

The same applies at higher multiplicities. For instance, the six-point (6-pt) partial amplitude again vanishes along a number of loci defined by the colour ordering. We will distinguish these based on the shape of their `causal diamond' in the kinematic mesh specific to that colour ordering \cite{Arkani-Hamed:2023swr}. At 6-pt, there are two possibilities, either a `skinny rectangle' of height one or a square of height two, see Fig. 
\begin{figure}
    \centering
    \includegraphics[width=0.4\linewidth]{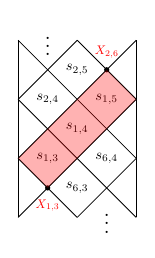}
     \includegraphics[width=0.4\linewidth]{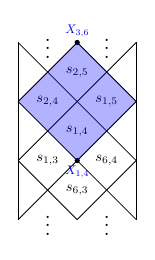}
    \caption{\em The skinny (red) and square (blue) causal diamonds illustrated in the 6-point kinematic mesh from \cite{Arkani-Hamed:2023swr}. The colored variables at the bottom and top of the causal diamonds are $X_B$ and $X_T$ mentioned later on in the text.}
    \label{fig:6pt mesh}
\end{figure}
Starting with the latter, the amplitude vanishes on the locus $s_{\alpha,\beta} = 0$ with
 \begin{align}
   \alpha = (1,2) \,, \quad \beta = (4,5) \,, \label{square}
 \end{align}
and similar for cyclic permutations. When only taking three equal to zero, the amplitude factorises and reads
 \begin{align}
  A_6^{\Phi^3}\to & \left(\frac{1}{X_{T}}+\frac{1}{X_{B}}\right) \times A^{\Phi^3, \, \rm{down}}_4 \times A^{\Phi^3, \, \rm{up}}_4\,,
  \label{Phi3 length-2 fac}
 \end{align}
and thus factorises into the product of two 4-pt amplitudes. $X_T$ and $X_B$ are the planar variables corresponding to the top and bottom vertices of the causal diamond; see \cite{Arkani-Hamed:2023swr}. For skinny rectangles, instead, the corresponding Mandelstams are $s_{\alpha, \beta}$ with
 \begin{align}
 \alpha = (1) \,, \quad \beta = (3,4,5) \,. \label{skinny}
 \end{align}
In order to vanish, all terms in the partial amplitude must be at least linear in one of these non-adjacent Mandelstam variables.
When taking all but one to zero, it factorises as
 \begin{align}
  A_6^{\Phi^3}   \to & (\frac{1}{X_{T}}+\frac{1}{X_{B}}) A^{\Phi^3}_3 A^{\Phi^3}_5\,,
  \label{Phi3 length-1 fac}
 \end{align}
into a product of three- and five-point amplitudes of the Tr$(\Phi^3)$ theory.  \\

\noindent
{\bf NLSM -} Secondly, in the absence of poles, the simplest 4-pt amplitude that vanishes at the locus is 
 \begin{align}
  A_4^{\rm NLSM} \sim u \,.
 \end{align}
This is the partial amplitude of the chiral NLSM, and provides the natural generalisation of \eqref{A4-Phi3} to contact interactions. 

Its higher-multiplicity amplitudes also vanish at the loci defined by the causal diamonds. Note that the vanishing of the partial amplitude at the skinny rectangle locus \eqref{skinny} foreshadows the Adler zero: in the absence of poles at $s_{1,2}$ or $s_{1,6}$ (guaranteed by the absence of cubic vertices - in contrast to the previous two cases), the vanishing of the partial amplitude at this locus implies the vanishing of the full amplitude in the soft limit $p_1 \to 0$. Thus the linearity in the soft Mandelstam variables of the skinny rectangle \eqref{skinny} goes beyond the Adler zero precisely by excluding adjacent soft Mandelstams. 

Moreover, the amplitudes of the NLSM factorise close to the loci of the causal diamonds as
 \begin{align}
    A^{\text{NLSM}}_6\to & \left(\frac{1}{X_B}+\frac{1}{X_T}\right) A_{4}^{\text{NLSM},\;{\rm up}}   A_{4}^{\text{NLSM},\;{\rm down}} \,, \notag \\
    A^{\text{NLSM}}_6 \to  & \left(X_T+X_B\right) A_3^{\Phi^3}  A_{5}^{\text{ext NLSM}}(\Phi,\pi,\pi,\Phi,\Phi) \,, 
 \end{align}
where 
 \begin{align}
    A_{5}^{\text{ext NLSM}}(\Phi,\pi,\pi,\Phi,\Phi)=& \frac{s_{2,3}+s_{3,4}}{s_{2,3,4}}-\frac{s_{3,5}}{s_{3,4,5}} \,.
\end{align}
The latter amplitude is generated by an extended theory that involves a novel \trphi sector \cite{Cachazo:2016njl},
\begin{align}
    {\cal L}^{\rm ext} = - \tfrac12 {\rm Tr} (D_{\mu}\Phi D^{\mu}\Phi) + {\rm Tr}(\Phi^3)  \,, \label{extension}
\end{align}
that contributes via minimal coupling to a composite connection that is defined in terms of pions. \\

\noindent
{\bf YMS - } 
The last theory whose partial amplitudes we would like to discuss has an additional flavour structure, and thus involves scalars $\phi^{aI}$. The 4-pt partial amplitude is 
 \begin{align}
  A^{\rm YMS}_4=  \frac{u}{s}\delta_{ab}\delta_{cd}  + \frac{u}{t}\delta_{ad}\delta_{bc}-\delta _{ac} \delta _{bd}\,. \label{A4-gluon}
 \end{align}
This structure arises from gluon exchange by the adjoint-valued scalars, as well as a quartic interaction that follows from the reduction of higher-dimensional Yang-Mills theory (and hence can be written in terms of scattering equations \cite{Cachazo:2014xea}). The above partial amplitude vanishes on the $u=0$ locus for the parts where the flavour structures is aligned with the colour ordering, i.e.~for $\delta_{ab} \delta_{cd}$ or $\delta_{bc} \delta_{ad}$.

At 6-pt, the partial amplitudes with flavour alignment (that is, proportional to $\delta_{ab} \delta_{cd} \delta_{ef}$ and $\delta_{af} \delta_{bc} \delta_{de}$) also vanish at exactly the same loci \eqref{square} and \eqref{skinny}. 
More generally, out of the 15 flavour structures of the YMS partial amplitudes, there are six structures that vanish for the skinny zeros defined by \eqref{skinny}: $\delta_{ab}\delta_{cd}\delta_{ef}$, $\delta_{af}\delta_{bc}\delta_{de}$, $\delta_{a b}\delta_{ce}\delta_{df}$, $\delta_{ab}\delta_{c f}\delta_{de}$, $\delta_{af}\delta_{be}\delta_{cd}$ and $\delta_{af}\delta_{bd}\delta_{ce}$. These are exactly those that do not connect the two sets of indices of \eqref{skinny}.

We therefore conjecture the following general rule for YMS amplitudes: 
\begin{quote} {\it For the partial amplitude with the alphabetic colour ordering, the flavour structures that do not pair up elements from $\alpha$ and $\beta$ vanish at the locus defined by these two sets of indices.}
\end{quote}
The resulting partial amplitude is therefore proportional to at least one $\delta_{\alpha_i \beta_j}$ on this locus. We have checked this conjecture up to and including 8-point.

Finally, in the factorisation limits defined by these loci, the different flavour structures split into two lower-point objects multiplied by a pre-factor. As a first example\footnote{We will comment on factorisation in other possible near-zero limits in section \ref{sec: color order zeros} and appendix \ref{app: all fact}.}, when setting all but $s_{1,5}$ in \eqref{square} to zero, the $\delta_{ab}\delta_{cd}\delta_{ef}$ part of the amplitude yields 
 \begin{align}
   A^{\text{YMS}}_6\to & \left(\frac{1}{X_B}+\frac{1}{X_T}\right) \times A_{4}^{\text{YMS},\;{\rm up}} \times A_{4}^{\text{YMS},\;{\rm down}} \nonumber\\
   = & \left(\frac{1}{s_{1,2,3}}+\frac{1}{s_{3,4,5}}\right)\times \frac{s_{1,2}+s_{2,3} }{s_{1,2}}\times \frac{s_{4,6}}{s_{4,5}+s_{4,6}}
   \label{YMS square 15}
   \end{align}
Secondly, setting all but $s_{1,5}$ in \eqref{skinny} to zero, this part of the amplitude reads
   \begin{align}
      A^{\text{YMS}}_6 \to & \frac{X_T+X_B}{X_B} \times A_3^{\Phi^3}\times A_{5}^{\text{ext YMS}}(\Phi,\phi,\phi,\Phi,\Phi) \,,
      \label{YMS s15 neq 0}
 \end{align}
where 
\begin{align}
    A_{5}^{\text{ext YMS}}(\Phi,\phi,\phi,\Phi,\Phi)=\frac{1}{s_{3,4}}\left(\frac{s_{2,3}+s_{3,4}}{s_{2,3,4}}-\frac{s_{3,5}}{s_{3,4,5}}\right) \,.
    \label{YMS s15 neq 0 5pt}
\end{align} 
In this case, the extension also involves a \trphi sector, 
\begin{align}
    {\cal L}^{\rm ext} = {\rm Tr}\left(-\tfrac12 D_{\mu}\Phi D^{\mu}\Phi + \Phi^3+\tfrac{g^2}{4}[\phi^I,\Phi][\phi^I,\Phi]\right) \,, \label{extension}
\end{align}
augmented with an additional quartic interaction.

%\rev{Note that we use $\phi$ to denote the flavoured and colored scalar in YMS and $\Phi$ for the colored scalar in ${\rm Tr}(\Phi^3)$.}

%\rev{One remark about the factorization of YMS is as follows. The YMS amplitude along $\delta_{ab}\delta_{cd}\delta_{ef}$ flavour structure factorizes into two lower-point objects no matter which variable is chosen to be non-zero. However, only when the right-most one is non-zero, can the lower-point objects be interpreted as amplitudes which are generated by \eqref{extension}.} \\

\section{Flavour pairing \& double zeros}
\label{sec: flav pairing double zeros}

\noindent
Going beyond colour-ordered partial amplitudes, we will demonstrate that scattering amplitudes can display multiple instance of the hidden zeros of \cite{Arkani-Hamed:2023swr}. As a first generalisation, we will retain some structure for the external scalars and require that they belong to a non-trivial group representation. Similar to the gluon exchange case, the full amplitudes at any multiplicity then consist of different `flavour structures' that pair up the external scalars via a product of Kronecker delta functions. The coefficients of these structures will be referred to as flavour-paired amplitudes. Without loss of generality, we will focus on the $\delta_{ab} \delta_{cd} \cdots$ flavour-paired amplitude in this section. \\

\noindent
{\bf GCS -} 
As a first step, consider 4-pt amplitude structures that vanish when imposing $t=0$ or $u=0$ separately. This leaves only the $s$-channel for exchange. The lowest order structure with two zeros is 
 \begin{align}
   A_4^{\rm GCS} \sim \frac{tu}{s} \,.
 \end{align}
This is the 4-pt amplitude of flavoured scalars coupled via graviton exchange. 

At 6-pt, the flavour-paired amplitude again vanishes at loci defined by causal diamonds of heights one and two. Starting with the latter squares, one such locus is \eqref{square}, while another (inequivalent) one is $\alpha=(1,2)$ and $\beta=(3,4)$. Other choices are related via index permutation that leave the flavour structure invariant. On all these flavour-compatible loci, indeed the part of the amplitude that is proportional to the above 'alphabetic' flavour pairing vanishes. 

Similarly, for the skinny rectangle \eqref{skinny} (all other choices being equivalent up to index permutations), the relevant part of the amplitude again vanishes. With for instance $\alpha=(1)$, this implies that $\beta$ can consist of any three out of the four indices $(3,4,5,6)$. In order to vanish on all these four loci, the flavour-paired amplitude has to be bilinear in this subset $s_{1\beta}$.

Near these zero loci, the amplitudes becomes the product of lower-point objects multiplied by a prefactor, exactly in the way described by the kinematic mesh; for instance, if we set $s_{1,4},s_{2,4},s_{2,5}=0$ and $s_{1,5}\neq 0$, the amplitude becomes
\begin{align}
  A_6^{\rm GCS}\to & \left(\frac{1}{s_{1,2,3}}+\frac{1}{s_{3,4,5}}\right)\times \frac{s_{2,3} \left(s_{1,2}+s_{2,3}\right)}{s_{1,2}}\frac{s_{4,5} s_{4,6}}{s_{4,5}+s_{4,6}}\nonumber\\
   = & \left(\frac{1}{X_B}+\frac{1}{X_T}\right)\times A_4^{\rm up,\;GCS}\times A_4^{\rm down,\;GCS}
   \label{GCS 2b2 ae neq 0}
\end{align}
The two 4-pt amplitudes are the pure scalar amplitudes with exchanged gluons. %, in the sense that $[AB][CD][EF]= [AB][CX]\times [XD][EF]$. 
Similarly, under the near-zero condition $s_{1,3},s_{1,4}=0,s_{1,5}\neq 0$, the amplitude factorizes into a product of 5-pt mixed amplitude and a prefactor 
\begin{align}
    A_6^{\rm GCS}  \to & \frac{X_T(X_B+X_T)}{X_B} A_3^{\Phi^3}  A_5^{{\rm ext \;GCS}}(\Phi,\rho,\rho,\Phi,\Phi) \,,
    \label{GCS s15 neq 0 5pt}
\end{align}
with
\begin{align}
     A_5^{{\rm ext\;GCS}}(\Phi,\rho,\rho,\Phi,\Phi)   = & \frac{s_{2,3}  \left(s_{2,3}+s_{3,4}\right)}{s_{3,4}s_{2,3,4}}+\frac{ s_{3,5} \left(s_{3,4}+s_{3,5}\right)}{s_{3,4}s_{3,4,5}}\nonumber\\
    &+\frac{ s_{3,6} \left(s_{3,4}+s_{3,6}\right)}{s_{3,4} s_{3,4,6}} \,.
    \label{mixed scaffolded GR}
\end{align}
The Lagrangian producing this mixed amplitude is provided by the extension
\begin{align}
    {\cal L}^{\text{ext}}=\sqrt{-g}( -\tfrac12 \partial^{\mu}\Phi\partial_{\mu}\Phi + \Phi^3) \,, \label{extension-metric}
\end{align}
where $g_{\mu\nu}=\eta_{\mu\nu} + h_{\mu\nu}$, and $\Phi$ is colorless. \\

%\rev{DELETE: The kinematics of this 5-pt amplitude and distribution of bi-flavoured scalars and $\Phi$  follows the pattern of \cite{Arkani-Hamed:2023swr}}.\\

\noindent
{\bf DBI -} 
Moving beyond exchange interactions with zeros at $t=0$ or $u=0$ for 4-pt, the lowest-order contact amplitude structure with this property is
 \begin{align}
  A^{\rm DBI}_4 \sim tu \,,
 \end{align}
which is the flavour amplitude of DBI. This theory realises a $(D+N)$-dimensional Poincar\'e algebra on $N$ scalars $\varphi^a$ in $D$ dimensions via non-linear transformations. Its central object is the induced metric
 \begin{align}
     g_{\mu\nu} = \eta_{\mu\nu} + \partial_\mu \varphi_a \partial_\nu \varphi^a \,, \label{DBI-met}
 \end{align}
which transforms covariantly under the higher-dimensional space-time symmetry.

Moving on to 6-pt, the DBI amplitude is a rather large expression. However, it is simple to verify that it again vanishes along a number of loci. These include the different squares and skinny rectangles that were already discussed for GCS; the part of the DBI-amplitude that is proportional to a particular flavour structure therefore also vanishes on all these loci. For the skinny rectangles, this again implies that these are bilinear in the soft Mandelstam variables. Moreover, DBI amplitudes are even, and therefore have no poles on single Mandelstams such as $s_{1,2}$. Soft bilinearity then implies that the full amplitude vanishes as 
 \begin{align}
     A_6^{\rm DBI} \sim p_1^2 \,,
 \end{align}
in the soft limit of $p_1 \to 0$. Therefore, similar to the single zero underlying the Adler zero of the NLSM, we find that the multiple loci of the DBI theory give rise to its extended soft limit.

Turning to factorisation, when only taking $s_{1,4}=s_{2,4}=s_{2,5}=0$, the amplitude factorises into 
 \begin{align}
  A_6^{\rm DBI}  \to & -\left(\frac{1}{s_{1,2,3}}+\frac{1}{s_{3,4,5}}\right) s_{2,3} \left(s_{1,2}+s_{2,3}\right) s_{4,5} s_{4,6} \nonumber \\
  = & \left(\frac{1}{X_B}+\frac{1}{X_T}\right)A_4^{{\rm up,DBI}}\times A_4^{{\rm down,DBI}}
  \label{DBI length-2 fac}
 \end{align}
in terms of two 4-pt DBI amplitudes. In the near-zero locus $s_{1,3},s_{1,4}\to 0$, the  DBI 6-pt flavour-paired amplitude factorizes into
\begin{align}
    A^{\rm DBI}_6 \to & X_T\left(X_B+X_T\right)A_3^{\Phi^3}A_5^{{\rm ext \;DBI}}(\Phi,\varphi,\varphi,\Phi,\Phi)\,,
     \label{DBI skinny fac}
\end{align}
with 
\begin{align}
  A_5^{{\rm ext \;DBI}}(\Phi,\varphi,\varphi,\Phi,\Phi) = & s_{3,4} A_5^{{\rm ext \;GCS}}(\Phi,\rho,\rho,\Phi,\Phi) \,.
  \label{DBI mixed}
    \end{align}
The Lagrangian of the mixed theory is given by the same form \eqref{extension-metric} but now with the DBI covariant metric \eqref{DBI-met} instead.

At this point we would like to comment on other factorizations. So far, we have shown that YMS, GCS and DBI have special factorization properties for one particular choice of the non-zero locus variable. For instance, for the square causal diamond, we have given the explicit expressions for $s_{14}=s_{24}=s_{25}=0$ and $s_{15}\neq 0$. We find that for other choices of the non-zero locus variable, the result is fully analogous and related by permutation. 

However, for skinny rectangles the sitation is slighly different. Above, we have discussed factorization for $s_{13}=s_{14}=0$ while the right-most Mandelstam $s_{15}\neq 0$. When keeping any of the other locus variables non-zero instead, the amplitudes still factorize but the lower-point objects\footnote{These lower-point objects can still be defined via CHY formalism; see \cite{Cao:2024qpp} for YMS.} are not necessarily related to the Lagrangian shown in this work. Full details of all possible factorizations for YMS, GCS and DBI are given in Appendix \ref{app: all fact}.

\section{Single scalar \& triple zeros}

\noindent
Finally, one can consider amplitudes that vanish when imposing $s=0$, $t=0$ or $u=0$ separately. This leaves us with no channels, and hence there can only be contact interactions. \\

\noindent
{\bf SG -} The lowest-order contact interaction with three zeros is 
 \begin{align}
  A_4^{\rm SG} \sim stu \,,
 \end{align}
which is the amplitude of the special Galileon. This is the unique single-scalar field theory that realises a non-linear symmetry that is quadratic in coordinates \cite{Hinterbichler} and has an interesting geometric interpretation \cite{Novotny, Roest}.

Given the absence of colour and flavour structures, the special Galileon 6-pt amplitude vanishes in any causal diamond spanned by a square, with 
 \begin{align}
   \alpha=(i,j) \,, \quad \beta=(k,l) \,,  \label{square-single}
 \end{align}
for any pair of pairs $(i,j),(k,l)$ (of which there are 15). Moreover, when only taking three out of the four equal to zero, {e.g.~} $s_{1,4},s_{2,4},s_{2,5}= 0$ (other choices being equivalent up to permutation), the amplitude becomes very simple and reads
 \begin{align}
     A_6^{\rm SG}  \to & \left(\frac{1}{s_{1,2,3}}+\frac{1}{s_{3,4,5}}\right) s_{1,2} s_{2,3} \left(s_{1,2}+s_{2,3}\right) s_{4,6} s_{4,5} \left(s_{4,6}+s_{4,6}\right) \nonumber \\
     = & \left(\frac{1}{X_B}+\frac{1}{X_T}\right)A_4^{{\rm up,SG}}\times A_4^{{\rm down,SG}} \,.
     \label{SG length-2 fac}
 \end{align}
This can be interpreted as a product of 4-pt SG amplitudes.

Looking at the skinny rectangles, the corresponding Mandelstams are an arbitrary selection of three variables from the soft Mandelstams with
 \begin{align}
   \alpha=(1) \,, \quad \beta=(\{2,3,4,5,6\}_{|3|}) \,.
   \label{skinny flavoured}
 \end{align}
The requirement that the amplitude must vanish at all these loci can only be satisfied by being trilinear in the above set. Combined with the evenness of the theory that ensures there are no odd poles, this implies that the amplitude scales with third power in the soft limit $p_1 \to 0$:
 \begin{align}
  A_6^{\rm SG} \sim p_1^3 \,.
 \end{align}
This was first found by Cheung et al \cite{Cheung:2014dqa} and later shown to follow from the non-linear symmetry \cite{Hinterbichler}. From the present perspective, we find however that the special Galileon has yet more special amplitudes: they vanish at a large number of kinematic loci.

At a near-zero locus, $s_{1,3},s_{1,4} = 0$ and $s_{1,5}\neq 0$, the amplitude becomes
\begin{align}
    A_6^{\rm SG}\to s_{1,5}s_{1,2} s_{1,6}\times A_3^{\Phi^3} A_5^{{\rm ext \;SG}}(\Phi,\varphi,\varphi,\Phi,\Phi) \,,
    \label{SG skinny 15}
\end{align}
where
\begin{align}
    A_5^{{\rm ext \;SG}}(\Phi,\varphi,\varphi,\Phi,\Phi)  = & s_{3,4}^2 A_5^{{\rm ext \; GCS}}(\Phi,\rho,\rho,\Phi,\Phi)  \,.
    \label{mixed SG}
\end{align}
This is exactly the mixed amplitude $A^{\rm ext \;SG}_5(\Phi,\varphi,\varphi,\Phi,\Phi)$ from \cite{Cachazo:2016njl} (up to permutation). It can also be generated from the extended Lagrangian \eqref{extension-metric} where the metric is given by
\begin{align}
    g_{\mu\nu} = \eta_{\mu\nu}+\partial_{\mu}\partial_{\rho}\varphi\partial^{\rho}\partial_{\nu}\varphi \, ,
    \label{SG-met}
\end{align}
being the special Galileon covariant metric.

\section{Hidden zeros and the double copy}

\noindent
We have seen in the previous sections that the single zeros of the colour-ordered theories of \cite{Arkani-Hamed:2023swr} can be extended to multiple zeros in three theories that lack colour ordering. In this section we will provide an explanation for this, by demonstrating that the hidden zeros are compatible with the double copy and thus are inherited from the colour-ordered theories.

The six theories identified above are very closely related to those arising in the double copy triangle of different scalar EFTs, with the triangle spanned by adjoint colour, fundamental flavour and singlet kinematic structures \cite{deNeeling:2022tsu}; for instance, the BCJ numerators \cite{Bern:2008qj,Bern:2010ue,Bern:2019prr,Bern:2022wqg} for flavour and kinematic structures at 4point\footnote{Extensions to higher derivatives can be found in e.g.~\cite{Carrasco:2022sck,Carrasco:2022lbm,Brown:2023srz,Li:2023wdm}.} are
 \begin{align}
  n^{\rm flav}_s & = \delta_{ab} \delta_{cd} (t-u) -  (\delta_{ac} \delta_{bd} - \delta_{bc} \delta_{ad}) s \,, \notag \\
  n^{\rm kin}_s & = s (t-u) \,.
 \end{align}
Together with colour, this leads to six possibilities that are closely related to the theories with zeros, as illustrated in figure 2: \begin{itemize}
    \item The coloured theories include the NLSM, as well as Yang-Mills-scalar (YMS) and the ${\rm Tr}(\Phi^3)$, which is a subsector of the bi-adjoint scalar (BAS).
    \item 
    Similarly, in the flavoured sector, one finds DBI, as well as gravity-coupled scalars. The latter is a subsector of the more general, bi-fundamental scalar theory that consists of a non-chiral $SO(M,N) / (SO(M) \times SO(N))$ NLSM coupled to gravity \cite{deNeeling:2022tsu}. Its scalar amplitudes coincide with those of the dimensional reduction of GR, referred to as Einstein-Maxwell-scalar (EMS) theory.
    \item 
    Finally, with only kinematics, the remaining possibility is the special Galileon, of the which the amplitudes is a function depending only on the Mandelstam variables. 
\end{itemize}
%Note that \rev{only some flavour components admit zero; for instance, the $\delta_{ab}\delta_{cd}\dots$ component of YMS has single zeros while the sum over all flavour components does not.}

\begin{figure}[t]
    \centering
   \includegraphics[scale=.90]{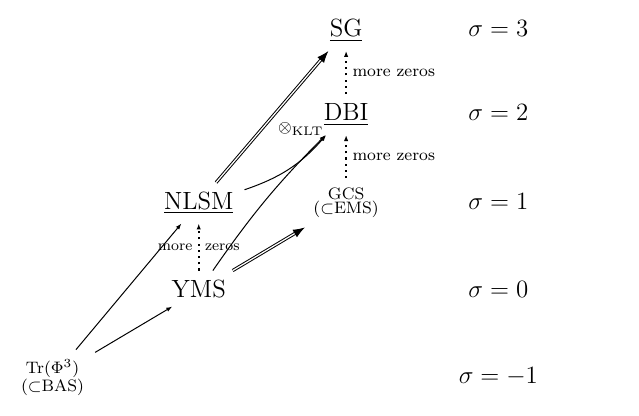}
    \caption{\it The different theories with zeros, arranged in terms of their soft degree $\sigma$. The underlined theories have no odd vertices and hence the zeros imply their soft behaviours. The vertical dotted lines indicate the increase in zeros for the DC-related theories.}
    \label{fig:tetrahedron}
\vspace{-0.25cm}
\end{figure}

Our approach to zeros and the DC will be based on the KLT formalism \cite{Kawai:1985xq} that combines two amplitudes $A_n$ and $\widetilde A_n$ into a third one $M_n$. In particular we will employ the formulation \cite{Bjerrum-Bohr:2010pnr}
\begin{align}
    M_n= & \sum_{\sigma\in S_{n-3}}\sum_{\substack{\alpha\in S_{i-1},\\ \beta\in S_{n-2-i}}}S[\alpha(\sigma(2,\dots,i))|\sigma(2,\dots,i)]_{k_1}\nonumber\\
    &\times S[\beta(\sigma(i+1,\dots,n-2))|\sigma(i+1,\dots,n-2)]_{k_{n-1}}\nonumber\\ 
    & \times A_n(1,\sigma(2,\dots,n-2),n-1,n) \nonumber \\
    & \times \widetilde{A}_n(\alpha(\sigma(2,\dots,i)),1,n-1,\beta(\sigma(i+1,\dots,n-2)),n) \,,
    \label{eq: general KLT formula}
\end{align}
where the KLT kernel is defined as
\begin{align}
    S[a_1,\dots,a_k|b_1\dots,b_k]_{p}=\prod_{t=1}^k \left( p\cdot k_{a_t}+\sum_{q>t}\theta(a_t,a_q)k_{a_t}\cdot k_{a_q}\right) \,, \label{KLT}
\end{align}
and $\theta(a_t,a_q)=0 (1)$ if the relative ordering of $a_t,a_q$ is the same (opposite) in $(a_1,\dots,a_k)$ and $(b_1,\dots,b_k)$. 

This formulation involves in general three sums over permutations: one 'long' permutation $\sigma$ and two shorter ones $\alpha, \beta$. The latter two can be seen to split the former one, with $i$ governing where the split takes place. This formulation is particularly suitable to show that the zeros of the $i \times (n-2-i)$ causal diamonds carry over from the single copy $\widetilde A$ to the double copy $M$. Without loss of generality, a causal diamond of height $i$ can be defined by all Mandelstam variables that link the {\it first set} $\{1, \ldots, i\}$ with the {\it second set} $\{i+1, \ldots, n-2 \}$ (while cyclicity generalises this to all causal diamonds of the same shape). At this locus, the above formula can be massaged into a more compact form by the following observation. 

The long permutation, that permutes all but $1$ of these, either respects the above separation into the two sets (and just permutes the indices within those two sets), or it permutes elements from one set with the other (on top of permutations within each set separately). In the latter case, the KLT matrix $S[\alpha|\sigma]_{k_1}$ vanishes. This can be seen as follows: consider in the permutation $\alpha(\sigma(2,\dots,j-1))$ the index of the second set that appears last. The corresponding factor in \eqref{KLT} then consists of a sum of Mandelstam variables that link the first and second sets, and hence vanish on the causal diamond. The long permutation $\sigma$ therefore splits up into two shorter permutations. The expression above then simplifies to
\begin{align}
    M_n= & \sum_{\substack{\alpha_{1,2} \in S_{i-1},\\ \beta_{1,2} \in S_{n-2-i}}} S[\alpha_1|\alpha_2]_{k_1} \times S[\beta_1 | \beta_2 ]_{k_{n-1}} \nonumber\\ 
    & \times A_n(1,\alpha_1,\beta_1,n-1,n) \times \widetilde{A}_n(\alpha_2,1,n-1,\beta_2,n) \,,
    \label{eq: general KLT formula2}
\end{align}
where $\alpha_{1,2}$ and $\beta_{1,2}$ are pairs of permutations of $(2,\ldots,i,)$ and $(i+1,\ldots,n-2)$, respectively.

In this form of the KLT formula, the remaining permutations of $\widetilde A$ do not affect this causal diamond; in other words, all partial amplitudes $\widetilde A$ that are summed over vanish at this causal diamond. These zeros therefore carry over from $\widetilde A$ to $M$, provided the other factor $A$ does not introduce any singularities at this locus. The latter is guaranteed when taken to be the NLSM on account of having only even amplitudes. This demonstrates that the SG inherits its hidden zeros from the NLSM.

The other possibility, with Yang-Mills-scalar as one of the factors, requires more discussion as it has singularities and flavour structures; at e.g.~4-pt, its partial amplitude is given by
 \begin{align}
  A_4(1,2,3,4) \sim u (\frac{\delta_{ab}\delta_{cd}}{s} + {\rm cyclic}) \,.
 \end{align}
which vanishes at $u=0$ for the two colour-compatible flavour structures $\delta_{ab}\delta_{cd}$ and $\delta_{bc}\delta_{ad}$ (and not for the third one). At higher points, we will assume the conjecture put forward in section \ref{sec: color order zeros} that, at a given (colour-compatible) locus, the only non-vanishing terms have flavour structures that link the two sets of indices (i.e.~are proportional to $\delta_{1 \beta_j}$ or $\delta_{\alpha_i \beta_j}$).

Under this assumption, it is straightforward to derive the zeros of DBI from the KLT formalism with $A^{\rm NLSM}$ and $\widetilde A^{\rm YMS}$: again all partial amplitudes $\widetilde A$ of the sum are only non-vanishing when the flavour structures link the first and second sets that define the locus. The resulting DBI amplitude therefore has the same property and hence vanishes along the complementary flavour structures.

The final possibility is to take both single copy factors to be Yang-Mills-scalar, with all external states being scalars. The resulting amplitudes are associated to a subsector of EMS where all external states are scalars with bi-fundamental indices, i.e.~$\rho^{a \bar a}$. For these amplitudes, it is convenient to define a trace notation for the linked bi-flavour structures. For example, at 4-pt, we define the short-hand notation for linked flavours as
\begin{align}
    [AB][CD]:=\delta_{ab}\delta_{\tilde{a}\tilde{b}}\delta_{cd}\delta_{\tilde{c}\tilde{d}}
    \label{4pt short trace}
\end{align}
while
\begin{align}
    [ABCD]:=\delta_{ab}\delta_{\tilde{b}\tilde{c}}\delta_{cd}\delta_{\tilde{a}\tilde{d}}
    \label{4pt long trace}
\end{align}
A bi-flavour structure is called a \emph{short trace} when the pairing of indices in both flavour and dual flavour coincides, such as \eqref{4pt short trace}. The gravitational interaction only gives `short traces', since the graviton does not carry any flavour, which effectively reduces to a single flavour structure\footnote{This is similar to the truncation from BAS to \trphi.}, and switches off the non-chiral NLSM. 

Under the KLT sum \eqref{eq: general KLT formula2}, the zeros of YMS carry over to the GCS amplitudes that do not pair up the first and second sets that define the locus. For example, at 6-point the flavour structures of YMS that do not pair up the two sets of skinny locus \eqref{skinny} are $\delta_{ab}\delta_{cd}\delta_{ef}$, $\delta_{af}\delta_{bc}\delta_{de}$, $\delta_{a b}\delta_{ce}\delta_{df}$, $\delta_{ab}\delta_{c f}\delta_{de}$, $\delta_{af}\delta_{be}\delta_{cd}$ and $\delta_{af}\delta_{bd}\delta_{ce}$. As long as both flavour and dual-flavour structures come from these six structures, the bi-flavoured GCS amplitude vanishes on this locus. This can be understood as follows.

%Since both single copies of GCS are YMS, there are two-particle poles in both single copies. One can track these two-particle poles by tracking the flavour delta's of YMS. In a YMS amplitude, a two-particle pole $1/s_{\alpha_1,\beta_1}$ must be accompanied by a $\delta_{\alpha_1,\beta_1}$, since this two-particle pole must come from a scalar-scalar-gluon minimal coupling, and the gluon does not carry any flavour. As it is mentioned in Sec.~\ref{sec: color order zeros}, any flavour structure of YMS vanishes on a zero locus as long as the flavour delta's never pair up indices from the two sets of indices $\alpha$ and $\beta$ associated to the locus. The reason is exactly that the zero locus variables are not associated to any two-particle pole of such flavour structure, hence do not trigger the blow-up of the YMS amplitudes.

In the KLT sum of GCS, any long permutation $\sigma$ can still be separated into two types: those that permute $\alpha$ and $\beta$ within themselves respectively, and those that permute $\alpha$ and $\beta$ into each other. The first type vanishes due to the same reason as before, that is, the zero locus is shared by all the color orderings where $\alpha$ and $\beta$ are permuted within themselves.  The second type vanishes as well, since the KLT kernel vanishes for the same argument, and the two-particle pole blow-up of neither single copy will be reached due to the constraints on the flavour structures. This ensures sure that the GCS amplitude with the required bi-flavour structure vanishes on the locus.

This completes our proof that the hidden zeros of single-copy theories (NLSM and YMS) carry over to double-copy theories (SG, DBI and the CGS subsector of EMS). This is contingent on the assumption made in the above, that YMS have the hidden zeros as put forward in the conjecture in section \ref{sec: color order zeros} (checked up to and including 8point).

\section{Conclusions}

\noindent
In this work, we have shown that three double copy theories (GCS, DBI and SG) have hidden zeros similar to the single copy theories. For DBI and the SG, these zeros underlie their well-known enhanced soft behaviours. Moreover, in the near-zero limits, the amplitudes of these theories factorize into lower-point objects. For causal diamonds such as the 6point square, these are amplitudes of the same theory; for the skinny rectange and when retaining only the rightmost variable, these are mixed amplitudes of an extended theories. In all three cases, the latter involves a covariantly coupled colorless $\Phi^3$, where the minimal coupling is to different metrics for the three theories. Finally, we prove that the hidden zeros of the double copy theories are inherited from the single copies via the most general version of the KLT formula.

Independently of the double copy relations, one could wonder whether the different causal diamonds of \cite{Arkani-Hamed:2023swr} correspond to independent zeros and factorisations, or whether they imply each other. Indeed, the zero of skinny rectangles combined with the near-zero factorization  imply the vanishing of height-two ($i=2)$ causal diamonds. To see this, mentally split the height-two causal diamond into two strips of length $n-4$. One of these can be seen as the near-zero limit of an $n$-point skinny rectangle, and hence implies the factorisation into an $(n-1)$-point amplitude. The second strip of the original causal diamond is then the full skinny rectangle for this $(n-1)$-point amplitude, which therefore has to vanish.

Unfortunately, there appears to be an obstacle to generalising this argument to $i=3$ and higher. When splitting these rectangles into smaller ones, the resulting shapes are $i-1$ Mandelstam variables short of defining a lower causal diamond. For $i \geq 3$, one therefore misses more than one Mandelstam variables and the near-zero factorisation of \cite{Arkani-Hamed:2023swr} does not apply. Possibly, this issue can be remedied by applying the new insights from \cite{Cao:2024gln}. In this work, it was found that the \trphi amplitude factorizes under a 2-split of Mandelstam variables, which is exactly one row shorter than the entire zero locus. This might allow for an extension of the previous argument to all $i$; we leave a further investigation for future work.

Another interesting topic for future research is to look for a possible stringy generalization of the special Galileon. The SG inherits all the zeros from the NLSM via the KLT double copy, which arises from the closed/open string relation. We have checked via some examples that the mod square generalization \cite{Arkani-Hamed:2019mrd} of the \trphi stringy amplitude also admits zeros. For GCS, the stringy integral is the mod square generalisation of that of YMS. There is no known UV completion or string theory embedding of the SG, although higher-derivative corrections up to finite ${\cal O}(\alpha')$ has been found \cite{Elvang:2018dco, Bartsch:2024amu}. It would be interesting to see whether the framework of surfaceology provides a new way to approach such a stringy generalization of the SG.

Finally, it is interesting to consider invariant polynomials that feature zero loci in a covariant manner as building blocks for amplitudes. Such a program will make manifest to what extent amplitudes are determined by their zero loci and pole structure.

\section*{Acknowledgements}

\noindent
We are grateful for stimulating discussions with Nichita Berzan, Tom\'a\v{s} Brauner, Scott Melville and Jasper Roosmale Nepveu, as well as the anonymous referee for insightful suggestions for improved presentation and further details.

\appendix

\section{Factorization for all choices of the non-zero locus variable}
\label{app: all fact}

\noindent
In this appendix, we show that, for the flavoured YMS, GCS and DBI theories, the near-zero factorization happens regardless of the choices of the non-zero locus variable, while the explicit forms of the lower-point amplitudes do depend on the choice. The discussion is based on the 6-point scenario. \\

\noindent
{\bf YMS}- The $\delta_{ab}\delta_{cd}\delta_{ef}$ flavour component of the YMS partial amplitude factorizes as follows:
\begin{itemize}
    \item[1)] Skinny rectangle: 
\begin{itemize}
    \item For $s_{1,3}\neq 0$ and $s_{1,4}=s_{1,5}=0$,
    \begin{align}
        A_6^{\rm YMS}\to \frac{ \left(s_{1,2}+s_{1,6}\right)}{s_{1,2}}\left(\frac{s_{3,4}+s_{4,5}}{s_{3,4}s_{3,4,5}}+\eqrev{\frac{s_{4,5}}{s_{3,4} s_{5,6}}}+\frac{s_{4,5}+s_{5,6}}{s_{5,6}s_{4,5,6}}\right)
        \label{YMS skinny 13}
    \end{align}
    \item For $s_{1,4}\neq 0$ and $s_{1,3}=s_{1,5}=0$, 
    \begin{align}
        A_6^{\rm YMS}\to -\frac{\left(s_{1,2}+s_{1,6}\right)}{s_{1,2}}\left(\eqrev{\frac{s_{3,5}}{s_{3,4}s_{5,6}}}+\frac{s_{3,5}}{s_{3,4}s_{3,4,5}}\right) 
        \label{YMS skinny 14}
    \end{align}
    \item For $s_{1,5}\neq 0$ and $s_{1,3}=s_{1,4}=0$, it is shown in \eqref{YMS s15 neq 0} and \eqref{YMS s15 neq 0 5pt}.
   \end{itemize}
   \item[2)] Square:
   \begin{itemize}
       \item For $s_{1,4}\neq 0$ and $s_{1,5}=s_{2,4}=s_{2,5}=0$,
       \begin{align}
           A_6^{\rm YMS}&\to  \left(\frac{1}{s_{1,2,3}}+\frac{1}{s_{3,4,5}}\right)\times \frac{s_{1,2}+s_{2,3}} {s_{1,2}}\times \frac{s_{4,5}+s_{5,6}}{s_{5,6}}
       \end{align}
        \item For $s_{1,5}\neq 0$ and $s_{1,4}=s_{2,4}=s_{2,5}=0$, the factorization is shown in \eqref{YMS square 15}.
        \item For $s_{2,4}\neq 0$ and $s_{1,4}=s_{1,5}=s_{2,5}=0$,
        \begin{align}
            A_6^{\rm YMS} \to -\left(\frac{1}{s_{1,2,3}}+\frac{1}{s_{3,4,5}}\right)\times\frac{s_{1,3} }{s_{1,2} }\times \frac{(s_{4,5}+s_{5,6})}{
            s_{5,6}}
        \end{align}
        \item For $s_{2,5}\neq 0$ and $s_{1,4}=s_{1,5}=s_{2,4}=0$.
        \begin{align}
             A_6^{\rm YMS} \to -\left(\frac{1}{s_{1,2,3}}+\frac{1}{s_{3,4,5}}\right)\times \frac{s_{1,3} }{s_{1,2} }\times \frac{s_{4,6}}{s_{4,5}+s_{4,6}}
        \end{align}
   \end{itemize}
\end{itemize} 

\noindent
{\bf GCS}- The $[AB][CD][EF]$ flavour component of GCS amplitude factorizes as follows
\begin{itemize}
    \item[1)] Skinny rectangle: 
\begin{itemize}
    \item For $s_{1,3}\neq 0$ and $s_{1,4}=s_{1,5}=0$,
    \begin{align}
        A_6^{\rm GCS}\to &\frac{s_{1,6} \left(s_{1,2}+s_{1,6}\right)}{s_{1,2}} \nonumber\\
        &\times s_{4,5} \left(\frac{s_{3,4}+s_{4,5}}{s_{3,4}s_{3,4,5}}+\eqrev{\frac{s_{4,5}}{s_{3,4} s_{5,6}}}+\frac{s_{4,5}+s_{5,6}}{s_{5,6}s_{4,5,6}}\right)
        \label{GCS skinny 13}
    \end{align}
    \item For $s_{1,4}\neq 0$ and $s_{1,3}=s_{1,5}=0$, 
    \begin{align}
        A_6^{\rm GCS}\to & \frac{s_{1,6} \left(s_{1,2}+s_{1,6}\right)}{s_{1,2}}  \nonumber\\
        &\times s_{3,5} \left(\eqrev{\frac{s_{3,5}}{s_{3,4} s_{5,6}}}+\frac{s_{3,4}+s_{3,5}}{s_{3,4} s_{3,4,5}}+\frac{s_{3,5}+s_{5,6}}{s_{5,6}s_{3,5,6}}\right)
        \label{GCS skinny 14}
    \end{align}
    \item For $s_{1,5}\neq 0$ and $s_{1,3}=s_{1,4}=0$, it is shown in \eqref{GCS s15 neq 0 5pt} and \eqref{mixed scaffolded GR}.
   \end{itemize}
   \item[2)] Square:
   \begin{itemize}
       \item For $s_{1,4}\neq 0$ and $s_{1,5}=s_{2,4}=s_{2,5}=0$,
       \begin{align}
           A_6^{\rm GCS}&\to \left(\frac{1}{s_{1,2,3}}+\frac{1}{s_{3,4,5}}\right)\frac{s_{2,3} \left(s_{1,2}+s_{2,3}\right)}{s_{1,2}}  \frac{s_{4,5}(s_{4,5}+s_{5,6})}{s_{5,6}}
       \end{align}
        \item For $s_{1,5}\neq 0$ and $s_{1,4}=s_{2,4}=s_{2,5}=0$, the factorization is shown in \eqref{GCS 2b2 ae neq 0}
        \item For $s_{2,4}\neq 0$ and $s_{1,4}=s_{1,5}=s_{2,5}=0$,
        \begin{align}
            A_6^{\rm GCS} \to \left(\frac{1}{s_{1,2,3}}+\frac{1}{s_{3,4,5}}\right)\frac{s_{2,3} \left(s_{1,2}+s_{2,3}\right) }{s_{1,2} }\frac{s_{4,5} (s_{4,5}+s_{5,6}) }{s_{5,6}}
        \end{align}
        \item For $s_{2,5}\neq 0$ and $s_{1,4}=s_{1,5}=s_{2,4}=0$,
        \begin{align}
            A_6^{\rm GCS} \to \left(\frac{1}{s_{1,2,3}}+\frac{1}{s_{3,4,5}}\right)\frac{s_{1,3} \left(s_{1,2}+s_{1,3}\right) }{s_{1,2} }\frac{ s_{4,5} s_{4,6}}{ s_{4,5}+s_{4,6}}
        \end{align}
   \end{itemize}
\end{itemize}

Note that for each of \eqref{YMS skinny 13}, \eqref{YMS skinny 14}, \eqref{GCS skinny 13}, and \eqref{GCS skinny 14}, some of the terms (indicated in red) cannot be generated from the extended Lagrangian. For example for \eqref{YMS skinny 13}, the red term has a numerator $s_{4,5}$, whose indices comes from two different outmost cubic vertices of a 5-point cubic diagram. To make such a term, some exotic operators such as $AA\Phi$ must be taken into account, which would break gauge invariance. Instead, as a non-Lagrangian approach, such `currents' (see \cite{Cao:2024qpp}) can be generated via the CHY formalism. \\

\noindent
{\bf DBI}- The $\delta_{ab}\delta_{cd}\delta_{ef}$ flavour component of DBI amplitude factorizes as follows
\begin{itemize}
    \item[1)] Skinny rectangle: 
\begin{itemize}
    \item For $s_{1,3}\neq 0$ and $s_{1,4}=s_{1,5}=0$,
    \begin{align}
        A_6^{\rm DBI}\to s_{1,6} \left(s_{1,2}+s_{1,6}\right)\times s_{4,5} \left(\frac{s_{3,4}+s_{4,5}}{s_{3,4,5}}-\eqrev{\frac{s_{4,6}}{s_{4,5,6}}}\right)
        \label{DBI skinny 13}
    \end{align}
    \item For $s_{1,4}\neq 0$ and $s_{1,3}=s_{1,5}=0$, 
    \begin{align}
        A_6^{\rm DBI}\to s_{1,6} \left(s_{1,2}+s_{1,6}\right)\times s_{3,5} \left(\frac{s_{3,4}+s_{3,5}}{s_{3,4,5}}-\eqrev{\frac{s_{3,6}}{s_{3,5,6}}}\right) 
        \label{DBI skinny 14}
    \end{align}
    \item For $s_{1,5}\neq 0$ and $s_{1,3}=s_{1,4}=0$, it is shown in \eqref{DBI skinny fac} and \eqref{DBI mixed}.
   \end{itemize}
   \item[2)] Square:
   \begin{itemize}
       \item For $s_{1,4}\neq 0$ and $s_{1,5}=s_{2,4}=s_{2,5}=0$,
       \begin{align}
           A_6^{\rm DBI}&\to  \left(\frac{1}{s_{1,2,3}}+\frac{1}{s_{3,4,5}}\right)\times s_{2,3} \left(s_{1,2}+s_{2,3}\right)  s_{4,5}(s_{4,5}+s_{5,6})
       \end{align}
        \item For $s_{1,5}\neq 0$ and $s_{1,4}=s_{2,4}=s_{2,5}=0$, the factorization is shown in \eqref{DBI length-2 fac}.
        \item For $s_{2,4}\neq 0$ and $s_{1,4}=s_{1,5}=s_{2,5}=0$,
        \begin{align}
            A_6^{\rm DBI} \to \left(\frac{1}{s_{1,2,3}}+\frac{1}{s_{3,4,5}}\right)\times s_{1,3} \left(s_{1,2}+s_{1,3}\right) s_{4,5} (s_{4,5}+s_{5,6})
        \end{align}
        \item For $s_{2,5}\neq 0$ and $s_{1,4}=s_{1,5}=s_{2,4}=0$.
        \begin{align}
             A_6^{\rm DBI} \to -\left(\frac{1}{s_{1,2,3}}+\frac{1}{s_{3,4,5}}\right)\times s_{1,3} \left(s_{1,2}+s_{1,3}\right) s_{4,5} s_{4,6}
        \end{align}
   \end{itemize}
\end{itemize}
Again, for \eqref{DBI skinny 13} and \eqref{DBI skinny 14}, the red terms cannot be generated by \eqref{DBI-met}. \\

\bibliography{ZEROS.bib}

\end{document}